\documentclass[11pt,a4paper,showpacs,showkeys,superscriptaddress]{article}

\usepackage{epsfig}
\usepackage{amsmath,amssymb}
\usepackage{graphicx}
\usepackage{xcolor}
\usepackage{subfigure}
\usepackage{amstext}
\usepackage{mathrsfs}

\usepackage{latexsym}
\usepackage{hyperref}
\hypersetup{colorlinks,%
  citecolor=blue,%
  linkcolor=red,%
  pdftex}

\usepackage[numbers,sort&compress,square]{natbib}

\begin{document}
\baselineskip 6mm

\newcommand{\nc}{\newcommand}
\newcommand{\rnc}{\renewcommand}



\newcommand{\tcb}{\textcolor{blue}}
\newcommand{\tcr}{\textcolor{red}}
\newcommand{\tcg}{\textcolor{green}}


\def\beq{\begin{equation}}
\def\eeq{\end{equation}}
\def\ba{\begin{array}}
\def\ea{\end{array}}
\def\bea{\begin{eqnarray}}
\def\eea{\end{eqnarray}}
\def\nn{\nonumber}


\def\CMP{Commun. Math. Phys.~}
\def\JHEP{JHEP~}
\def\Pre{Preprint}
\def\PRL{Phys. Rev. Lett.~}
\def\PR {Phys. Rev.~}
\def\CQG {Class. Quant. Grav.~}
\def\PL {Phys. Lett.~}
\def\NP {Nucl. Phys.~}

\def\G{\Gamma}

\def\S{{\bf S}}
\def\C{{\bf C}}
\def\Z{{\bf Z}}
\def\R{{\bf R}}
\def\N{{\bf N}}
\def\M{{\bf M}}
\def\P{{\bf P}}
\def\bm{{\bf m}}
\def\bn{{\bf n}}

\def\CA{{\cal A}}
\def\CB{{\cal B}}
\def\CC{{\cal C}}
\def\CD{{\cal D}}
\def\CE{{\cal E}}
\def\CF{{\cal F}}
\def\CH{{\cal H}}
\def\CM{{\cal M}}
\def\CG{{\cal G}}
\def\CI{{\cal I}}
\def\CJ{{\cal J}}
\def\CL{{\cal L}}
\def\CK{{\cal K}}
\def\CN{{\cal N}}
\def\CO{{\cal O}}
\def\CP{{\cal P}}
\def\CQ{{\cal Q}}
\def\CR{{\cal R}}
\def\CS{{\cal S}}
\def\CT{{\cal T}}
\def\CU{{\cal U}}
\def\CV{{\cal V}}
\def\CW{{\cal W}}
\def\CX{{\cal X}}
\def\CY{{\cal Y}}
\def\CZ{{\cal Z}}

\def\We{{W_{\mbox{eff}}}}


\newcommand{\Lie}{\pounds}

\newcommand{\p}{\partial}
\newcommand{\bp}{\bar{\partial}}

\newcommand{\half}{\frac{1}{2}}

\newcommand{\bfalpha}{{\mbox{\boldmath $\alpha$}}}
\newcommand{\bfbeta}{{\mbox{\boldmath $\beta$}}}
\newcommand{\bfgamma}{{\mbox{\boldmath $\gamma$}}}
\newcommand{\bfmu}{{\mbox{\boldmath $\mu$}}}
\newcommand{\bfpi}{{\mbox{\boldmath $\pi$}}}
\newcommand{\bfvarpi}{{\mbox{\boldmath $\varpi$}}}
\newcommand{\bftau}{{\mbox{\boldmath $\tau$}}}
\newcommand{\bfeta}{{\mbox{\boldmath $\eta$}}}
\newcommand{\bfxi}{{\mbox{\boldmath $\xi$}}}
\newcommand{\bfkappa}{{\mbox{\boldmath $\kappa$}}}
\newcommand{\bfepsilon}{{\mbox{\boldmath $\epsilon$}}}
\newcommand{\bfTheta}{{\mbox{\boldmath $\Theta$}}}

\newcommand{\bz}{{\bar{z}}}

\newcommand{\dalpha}{\dot{\alpha}}
\newcommand{\dbeta}{\dot{\beta}}
\newcommand{\blambda}{\bar{\lambda}}
\newcommand{\btheta}{{\bar{\theta}}}
\newcommand{\bsigma}{{{\bar{\sigma}}}}
\newcommand{\bepsilon}{{\bar{\epsilon}}}
\newcommand{\bpsi}{{\bar{\psi}}}


\def\ct{\cite}
\def\la{\label}
\def\eq#1{(\ref{#1})}


\def\a{\alpha}
\def\b{\beta}
\def\g{\gamma}
\def\G{\Gamma}
\def\d{\delta}
\def\D{\Delta}
\def\ep{\epsilon}
\def\e{\eta}
\def\ph{\phi}
\def\Ph{\Phi}
\def\ps{\psi}
\def\Ps{\Psi}
\def\k{\kappa}
\def\l{\lambda}
\def\L{\Lambda}
\def\m{\mu}
\def\n{\nu}
\def\th{\theta}
\def\Th{\Theta}
\def\r{\rho}
\def\s{\sigma}
\def\S{\Sigma}
\def\ta{\tau}
\def\o{\omega}
\def\O{\Omega}
\def\pr{\prime}
\def\f{\varphi}


\def\half{\frac{1}{2}}

\def\goto{\rightarrow}

\def\na{\nabla}
\def\grad{\nabla}
\def\curl{\nabla\times}
\def\div{\nabla\cdot}
\def\pa{\partial}

\def\bra{\left\langle}
\def\ket{\right\rangle}
\def\lb{\left[}
\def\lc{\left\{}
\def\ls{\left(}
\def\lp{\left.}
\def\rp{\right.}
\def\rb{\right]}
\def\rc{\right\}}
\def\rs{\right)}
\def\cl{\mathcal{l}}

\def\vac#1{\mid #1 \rangle}

\def\td#1{\tilde{#1}}
\def\check{ \maltese {\bf Check!}}


\def\Tr{{\rm Tr}\,}
\def\det{{\rm det}\,}


\def\bc#1{\nnindent {\bf $\bullet$ #1} \\ }
\def\ch {$<Check!>$ }
\def\ss {\vspace{1.5cm}}

\begin{titlepage}
%
%
%
%
%
%
%
%
\begin{center}
{\Large \bf Thermodynamic Volume and the Extended Smarr Relation}
%
\vskip 1. cm
  { Seungjoon Hyun\footnote{e-mail : sjhyun@yonsei.ac.kr}, Jaehoon Jeong\footnote{e-mail : j.jeong@yonsei.ac.kr}, Sang-A Park\footnote{e-mail : sangapark@yonsei.ac.kr},
  Sang-Heon Yi\footnote{e-mail : shyi@yonsei.ac.kr} 
 }
\vskip 0.5cm
{\it Department of Physics, College of Science, Yonsei University, Seoul 120-749, Korea}\\
\end{center}
\thispagestyle{empty}
\vskip1.5cm
%
%
\centerline{\bf ABSTRACT} \vskip 4mm 
 \vspace{1cm} 
\noindent   
We continue to explore the scaling transformation in the reduced action formalism of  gravity models. As an extension of our construction, we consider  the extended forms of the Smarr relation for various black holes, adopting the cosmological constant as the bulk pressure as in some literatures on black holes.  Firstly, by using the quasi-local formalism for charges, we show that, in a general theory of gravity, the volume in the black hole thermodynamics could be defined as  the thermodynamic conjugate variable to the bulk pressure in such a way that  the first law can be extended  consistently. This, so called, thermodynamic volume can be expressed explicitly in terms of the metric and field variables. Then, by using the scaling transformation allowed in the reduced action formulation, we obtain the extended Smarr relation involving the bulk pressure and the thermodynamic volume.  In our approach, we do not resort to Euler's homogeneous scaling  of charges    while incorporating the would-be hairy contribution without any difficulty. 
\vspace{2cm} 
%
%
\end{titlepage}
\renewcommand{\thefootnote}{\arabic{footnote}}
\setcounter{footnote}{0}
%
%
%
%

\section{Introduction}
Recently, it has been recognized that the cosmological constant can be treated as the bulk pressure in the black hole thermodynamics. This approach opens up a new perspective for black hole thermodynamics (See for a recent review~\cite{Kubiznak:2016qmn}). This adoption of the cosmological constant as one of the thermodynamic variables extends the first law as well as  endows a meaning to the Smarr relation on the AdS black holes. Historically, the first law of black hole thermodynamics~\cite{Bekenstein:1971hc,Hawking:1974sw,Hawking:1974rv,Bardeen:1973gs} is granted as universal but the Smarr(-Gibbs-Duhem) relation~\cite{Smarr:1972kt} is regarded as one of the particular properties of specific black holes.  Concretely, in contrast to the first law,  the coefficients of terms in the Smarr relation depend on the spacetime dimension and  the relation itself is regarded as nonexistent in the asymptotically AdS space (see for instance~\cite{Gibbons:2004ai}). However, by treating the cosmological constant as a thermodynamic variable, it has been shown that the first law and the Smarr relation could be formulated uniformly~\cite{Henneaux:1984ji,Teitelboim:1985dp,Caldarelli:1999xj,Wang:2006eb,Sekiwa:2006qj,LarranagaRubio:2007ut,Kastor:2009wy,Dolan:2010ha,Cvetic:2010jb,Azreg-Ainou:2014lua,Azreg-Ainou:2014twa,Couch:2016exn}. In this approach, the cosmological constant is identified as the bulk pressure and then, its conjugate thermodynamic variable is as the volume. 

Though the adoption of the cosmological constant as a thermodynamic variable may be bizarre from the perspective that  it is one of  Lagrangian parameters, which is usually fixed, not the parameter in the solutions of the equations of motion(EOM), it has led us to an interesting uniform description of the first law and the Smarr relations  for various asymptotic geometries including the asymptotic AdS and Lifshitz spacetimes. Concretely speaking, after the recognition of the cosmological constant as a thermodynamic variable, it turns out that the Smarr relation is a simple consequence of the Euler's homogeneous scaling property of the various thermodynamic quantities. 

However, one may note  that in the derivation of the Smarr relation one needs to assume the definite  on-shell  scaling behavior of the various thermodynamic quantities with respect to the cosmological  constant. This assumption could be valid but seems to be non manifest especially for hairy black holes. As has been well known,  scalar hairy black holes are admitted in the asymptotic AdS spacetime, and the hairy parameters could appear non-linearly in these black hole solutions. Since the hairy effects could enter non-linearly, their behavior under the scaling would be non-homogeneous.   Therefore, it is unclear how to implement the hairy parameters in the Euler's homogeneous scaling arguments and it would be worth to pursue  another approach toward the Smarr relation by treating the cosmological constant as a thermodynamic variable.  

On the other hand, the off-shell scaling transformation in the reduced action formalism is successfully utilized to obtain the Smarr relation in a rather uniform fashion~\cite{Hyun:2015tia,Ahn:2015uza,Ahn:2015shg,Hyun:2016isn}. When this off-shell scaling transformation is a symmetry of the reduced action, one can obtain the conventional Smarr relation for planar black holes. 
Though planar black holes are invoked for this scaling transformation,  it is anticipated that  the scaling transformation method could reproduce or extend the  incorporation of the cosmological constant  as a thermodynamic variable. Even for the non-planar black holes, one may use the off-shell scaling transformation, which is not a symmetry, for the extended thermodynamic relations. 

In section 2, we use the quasi-local formalism for conserved charges to find  the thermodynamic volume as the conjugate to the bulk pressure, under which the first law of black hole thermodynamics can be extended consistently. It turns out that the thermodynamic volume  can be expressed in terms of the metric and field variables. In section 3, we
extend our recent off-shell scaling method to cover  the extended Smarr relation in various black holes with the  cosmological constant as a thermodynamic variable. The off-shell nature of the scaling transformation of various fields in the reduced action is essential to obtain the extended Smarr relation, which includes the previously known results as a special case.  In section 4, we give several examples including AdS black holes with the scalar hair and Lifshitz black holes.

\section{The quasi-local approach to the extended first law}

In this section, we give the arguments for  the extended first law of black hole thermodynamics with the variation of the parameters in the Lagrangian or the equations of motion, which would be a straightforward extension of the Einstein gravity results in the literatures to those in a generic theory gravity. 

For definiteness, we focus on the case of the  cosmological constant, $\Lambda$. However, we would like to emphasize that the following procedure could be applied to any other Lagrangian parameter. First, let us introduce the extended  variation  $\tilde{\delta} \equiv \delta +{\delta}_{\Lambda}$.
This extended variation  is composed of the conventional field variation, $\delta$, accompanied by the variation with respect to the cosmological constant, ${\delta}_{\Lambda}$, which should be kept constant in the conventional variation. Note also that the variation with respect to the cosmological constant $\Lambda$ could be  further decomposed into parts from the variation of the field with respect to the cosmological constant and from the explicit dependence on the cosmological constant. Explicitly, the variation of  a function  $F(\Psi,\Lambda)$ with respect to the cosmological constant could be written as
\begin{equation} \label{}
{\delta}_{\Lambda}F(\Psi, \Lambda) = \frac{\delta F}{\delta \Psi}\delta_{\Lambda}\Psi + \frac{\p F}{\p \Lambda}\delta \Lambda\,,
\end{equation}
 where $\Psi$ denotes collectively the metric $g_{\mu\nu}$ and matter fields $\psi$.
Just like the conventional linearization, the linearization of the solution $\tilde{\delta} \Psi$ in the extended sense would satisfy the linearized equations of motions of metric and matter fields, $\tilde{\delta}{\cal E}_{\mu\nu}=0$ and $\tilde{\delta}{\cal E}_{\psi}=0$, respectively.

In the following, we would like to obtain the extended first law of black hole thermodynamics from the quasi-local Abbott-Deser-Tekin(ADT) formalism~\cite{Abbott:1981ff,Abbott:1982jh,Deser:2002rt,Kim:2013zha,Kim:2013cor,Hyun:2014sha,Hyun:2014nma,Hyun:2016dvt}. 
Recall that the off-shell conserved ADT current for a Killing vector $\xi$ could  be introduced as 
\begin{equation} \label{ADTcurr}
\sqrt{-g}{\cal J}^{\mu}_{ADT}(\xi\,;\, \Psi,  \delta\Psi) = \delta\Big(\sqrt{-g}{\bf E}^{\mu}_{~\nu}\xi^{\nu}\Big) + \frac{1}{2}\sqrt{-g}{\cal E}_{\Psi} \delta \Psi\,,
\end{equation}
where we have set $\delta\xi=0$ for simplicity.  And then, one could introduce the ADT potential as
\begin{equation} \label{ADTpot}
\sqrt{-g}{\cal J}^{\mu}_{ADT}(\xi\,;\, \Psi,  \delta\Psi)  = \p_{\nu}\Big(\sqrt{-g}Q^{\mu\nu}_{ADT}(\xi\,;\, \Psi,  \delta\Psi) \Big)\,.
\end{equation}
Let us also recall that the infinitesimal expression of the charge for a Killing vector $\xi$ is given through the ADT potential $Q^{\mu\nu}_{ADT}$ as (See~\cite{Hyun:2014sha,Hyun:2016dvt} for our conventions.)
\begin{equation} \label{}
 \delta Q_{ADT}(\xi)  = \frac{1}{8\pi G}  \int d^{D-2}x_{\mu\nu} \sqrt{-g}Q^{\mu\nu}_{ADT}(\xi\,;\, \Psi,  \delta \Psi) \,.
\end{equation}
By the integration along the one-parameter path in the solutions space, the finite expression of charges for the Killing vector $\xi$ is given by
\begin{equation} \label{}
Q_{ADT}(\xi) = \frac{1}{8\pi G} \int_{0}^{1} ds  \int d^{D-2}x_{\mu\nu} \sqrt{-g}Q^{\mu\nu}_{ADT}(\xi\,;\, \Psi,  \delta_{s} \Psi) \,,
\end{equation}
which could also be written in term of the Noether potential, $K$ and the surface term $\Theta$ as
\begin{equation} \label{}
Q_{ADT}(\xi) = \frac{1}{16\pi G}  \int d^{D-2}x_{\mu\nu}\Big[\Delta K^{\mu\nu}(\xi\,;\, \Psi,  \delta_{s} \Psi) - \int^{1}_{0} ds~2 \xi^{[\mu}\Theta^{\nu]}(\Psi, \delta_{s}\Psi)\Big] \,,
\end{equation}
where $\Delta K$ denotes the difference between the black hole configuration and the background, $\Delta K = K(s=1)-K(s=0)$. 

Now, we would like to consider the  off-shell ADT current and ADT potential  for the varying cosmological constant. To this purpose, let us consider the extended variation of ${\bf E}^{\m}_{~\nu}$, first.  Because of the explicit dependence of the Lagrangian on the cosmological constant, or the dependence of the expression ${\bf E}^{\mu}_{~\nu}(\Psi,\Lambda)$ on the cosmological constant, one could set
\begin{equation} \label{extvarEOM}
\tilde{\delta}{\bf E}^{\mu}_{~\nu}= \frac{\delta {\bf E}^{\mu}_{~\nu}}{\delta \Psi}\tilde{\delta} \Psi 
+  \mathscr{E}^{\mu}_{~\nu} \delta \Lambda\,,
\end{equation}
where $\mathscr{E}^{\mu}_{~\nu} (\Psi, \Lambda)$ is defined by $ \mathscr{E}^{\mu}_{~\nu} \equiv \frac{\p {\bf E}^{\mu}_{~\nu}}{\p \Lambda}$. 
%
%
By assuming that the additional part $\mathscr{E}_{\mu\nu} (\Psi, \Lambda)$ in the extended variation of ${\bf E}^{\mu}_{~\nu}$ is covariantly conserved, one can also introduce the potential for the additional part  as 
\begin{equation} \label{Ompot}
\mathscr{E}^{\mu}_{~\nu} (\Psi, \Lambda)\xi^{\nu} = \nabla_{\nu}\Omega^{\mu\nu} (\xi\,;\, \Psi, \Lambda)\,,
\end{equation}
where $\Omega^{\mu\nu}$ is an anti-symmetric tensor. 
At this stage, it would be useful to recall that both the ADT potential $Q^{\mu\nu}_{ADT}$ and the potential $\Omega^{\mu\nu}$ have the ambiguities up to the total derivatives by construction.  
Note that the ADT current has the vanishing on-shell property, {\it i.e.} 
\[   
{\cal J}^{\mu}_{ADT}(\xi\,;\, \Psi,\delta \Psi\,|\Lambda)\Big|_{on-shell}=0\,,
\]
and that the extended variation of ${\bf E}^{\mu}_{~\nu}$ also has the same property.  Through the definition of the ADT current in Eq.~(\ref{ADTcurr}) under the on-shell condition, one can see that the ADT current for the varying cosmological constant   becomes 
\begin{align} \label{CurrentRel}
	\Big[{\cal J}^{\mu}_{ADT}(\xi\,;\, \Psi, \delta_{\Lambda} \Psi\,|\, \Lambda)+   \mathscr{E}^{\mu}_{~\nu}(\xi\,;\, \Psi, \Lambda)\xi^{\nu}~ \delta \Lambda\Big]_{on-shell} =0\,.
\end{align}

One may introduce the infinitesimal expression of the charge for a Killing vector $\xi$  in the context of the extended variation, just like  the above ADT potential $Q^{\mu\nu}_{ADT}$, as 
\begin{align}   \label{extADTcharge}
 \tilde{\delta} Q_{ADT}(\xi) &= \delta Q_{ADT}(\xi)  + \delta_{\Lambda}  Q_{ADT}(\xi)\,,
\end{align}
where one may  define $\delta_{\Lambda}Q_{ADT}$ as 
\begin{equation} \label{}
\delta_{\Lambda}Q_{ADT}(\xi) \equiv  \delta_{\Lambda}\bigg[\frac{1}{8\pi G} \int ds  \int d^{D-2}x_{\mu\nu} \sqrt{-g}Q^{\mu\nu}_{ADT}(\xi\,;\, \Psi,  \delta_{s} \Psi\,|\, \Lambda) \bigg]\,,
\end{equation}
which could also be written as
\begin{equation} \label{finiteQvar}
\delta_{\Lambda}Q_{ADT}(\xi) = \delta_{\Lambda}\bigg[ \frac{1}{16\pi G}  \int d^{D-2}x_{\mu\nu}\Big(\Delta K^{\mu\nu}(\xi\,;\, \Psi,  \delta_{s} \Psi) - \int ds~2 \xi^{[\mu}\Theta^{\nu]}(\Psi, \delta_{s}\Psi)\Big) \bigg] \,.
\end{equation}
%
%
%
%
Here, we should be  careful in defining the charge,  since the variation of the cosmological constant could affect the background and then the change of the background could enter in the charge expression. This background changing effect is not the case we are trying to formulate. Our aim is to  construct the charge expression which should be related to the black hole properties, not those of the background spacetime. 
Therefore, we need to determine how to compare the charges of black hole solutions among theories with different cosmological constants. 
Basically, in our setup, we are trying to vary the cosmological constant in the charge expression, after we obtain the charges of black holes by a  conventional method. 
%

Now, we would like to rewrite the above expression of $\delta_{\Lambda}Q_{ADT}$ in terms of the ADT potential $Q^{\mu\nu}_{ADT}(\xi\,;\, \Psi,  \delta_{\Lambda} \Psi\,|\, \Lambda)$.  It is straightforward to see that the Noether potential part satisfies $(\delta\delta_{\Lambda} - \delta_{\Lambda}\delta) K =0$. One may note that the unwanted background dependent contribution from the varying cosmological constant comes from the surface term $\Theta$ and the part of $K(s=0)$. In general, the surface term satisfies 
\begin{equation} \label{}
\delta_{\Lambda}\Theta^{\mu}(\delta \Psi)  - \delta \Theta^{\mu}(\delta_{\Lambda}\Psi) = \omega^{\mu}(\delta_{\Lambda}\Psi, \delta\Psi)\,,
\end{equation}
where $\omega^{\mu}$ denotes the, so-called, symplectic current~\cite{Lee:1990nz}. 
Recall that the given black hole configuration corresponds to $s=1$ and the background does to $s=0$ and that 
\[   
\delta_{\Lambda}K^{\mu\nu} =  \delta_{\Lambda}\Delta K^{\mu\nu} + \delta_{\Lambda}K^{\mu\nu}_{s=0}\,, \qquad  \Theta^{\mu}(\Psi,\d_{\Lambda} \Psi) = \int^{1}_{0} ds  ~\Big[\delta_{s}\Theta^{\mu}(\Psi,\d_{\Lambda} \Psi)\Big] + \Theta^{\mu}(\Psi,\d_{\Lambda} \Psi)\big|_{s=0}\,.
\]

Rewriting the above expression of $\delta_{\Lambda}Q_{ADT}$ in Eq.(\ref{finiteQvar}) in terms of $Q^{\mu\nu}_{ADT}(\xi\,;\, \Psi,  \delta_{\Lambda} \Psi\,|\, \Lambda)$ and the symplectic current $\omega^{\mu}$, one can see that
\begin{align}   \label{exQADT}
\delta_{\Lambda} Q_{ADT}(\xi) 
& = \frac{1}{8\pi G}  \int d^{D-2}x_{\mu\nu} \bigg[ \sqrt{-g}  Q^{\mu\nu}_{ADT}(\xi\,;\, \Psi, \delta_{\Lambda}\Psi\,|\,\Lambda) \bigg. \nn\\
&\hspace{2.4cm} \bigg. - \frac{1}{2} \Big[ \delta_{\Lambda}K^{\mu\nu}- 2\xi^{[\mu}\Theta^{\nu]}(\Psi,\d_{\Lambda} \Psi) \Big]_{s=0} -   \int ds~  \xi^{[\mu}\omega^{\nu]}(\delta_{\Lambda}\Psi,\delta_{s}\Psi)\bigg]   \nn \\
&\equiv  \frac{1}{8\pi G} \int   d^{D-2}x_{\mu\nu}~   \Delta'\Big[\sqrt{-g}  Q^{\mu\nu}_{ADT}(\xi\,;\, \Psi, \delta_{\Lambda}\Psi\,|\, \Lambda ) \Big]\,, 
\end{align}
where $\Delta'$ denotes the subtraction by the background variation of Noether potential, $K$, and the symplectic current. In the following, we  call the contribution by $K^{\mu\nu}_{s=0}$, $\Theta^{\mu}_{s=0}$, and $\omega^{\mu}(\delta_{\Lambda}\Psi,\delta \Psi)$ as the `background contribution'. 

%
%

As equipped with the above construction, let us consider the extended first law.  The conventional first law of AdS-Kerr black holes would be written in terms of the conventional on-shell variations as
\begin{equation} \label{}
\delta M = T_{H}\delta {\cal S}_{GH} + \Omega_{H}\delta J\,,
\end{equation}
which holds  by regarding the cosmological constant as a fixed parameter.
%
%
%
%
By allowing the variation of the cosmological constant in  the charge expressions of the same black holes, the extended first law would be written in terms of the extended variation $\tilde{\delta}$ as 
\[   
\tilde{\delta} M = T_{H}\tilde{\delta} {\cal S}_{BH} +  \Omega_{H}\, \tilde{\delta} J + \Big[\delta_{\Lambda}M -  \Omega_{H}\, \delta_{\Lambda}  J  - T_{H}\, \delta_{\Lambda} {\cal S}_{BH}\Big]\,,
\]
where we have used the relation among the variations of charges in Eq.~(\ref{extADTcharge}). 
Note that the bracket part in the right hand side of the above equation comes solely  from the variation of the cosmological constant. Hence, by representing the last bracket part  in the form of  
\begin{equation} \label{Step1}
\delta_{\Lambda}M - \Omega_{H}\, \delta_{\Lambda}  J   - T_{H}\, \delta_{\Lambda} {\cal S}_{BH} = -\frac{ V}{8\pi G}~ \delta \Lambda \,,
\end{equation}
we introduce  the so-called thermodynamic volume $V$. At this stage, the thermodynamic volume is introduced just as the coefficient of the variation of the cosmological constant, $\delta\Lambda$. Finally the extended first law could be written as
\begin{equation} \label{extFirstLaw}
\tilde{\delta} M = T_{H}\tilde{\delta} {\cal S}_{BH} +  \Omega_{H}\, \tilde{\delta} J +  V\tilde{\delta} P\,,
\end{equation}
where the pressure is defined by $P\equiv -\Lambda/8\pi G$ and thus $\tilde{\delta} P = -\delta \Lambda/8\pi G$.
We would like to emphasize that this extended first law holds even with scalar hairs, since all those contributions are included in the above derivation. See some related discussions given in \cite{Hyun:2016isn}.

In order to obtain the explicit form of the volume $V$, we proceed as follows. 
Firstly,  recall that Eq.~(\ref{CurrentRel})  implies that
\[   
 \frac{1}{8\pi G} \Big(\int_{\infty}-\int_{\cal B} \Big)   d^{D-2}x_{\mu\nu}~    \sqrt{-g}\Big[ Q^{\mu\nu}_{ADT}(\xi_{H}\,;\, \Psi, \delta_{\Lambda} \Psi\,|\, \Lambda) + \Omega^{\mu\nu} (\xi\,;\, \Psi, \Lambda) \delta\Lambda\Big] =0\,.
\]
Secondly, by 
using  the charge expression written in terms of the ADT  potential $Q^{\mu\nu}_{ADT}$, one can see that
\begin{align}   \label{Step2}
 & \delta_{\Lambda}M - \Omega_{H}\, \delta_{\Lambda}  J   - T_{H}\, \delta_{\Lambda} {\cal S}_{BH}  \\
   &\qquad = \frac{1}{8\pi G} \Big( \int_{\infty} - \int_{\cal B} \Big)  d^{D-2}x_{\mu\nu}~   \Delta'\Big[\sqrt{-g}  Q^{\mu\nu}_{ADT}(\xi\,;\, \Psi, \delta_{\Lambda}\Psi\,|\, \Lambda ) \Big]   \nn \\
 &\qquad = -  \frac{1}{8\pi G} \Big( \int_{\infty} - \int_{\cal B} \Big)  d^{D-2}x_{\mu\nu}~    \Delta'\Big[\sqrt{-g}\Omega^{\mu\nu} (\xi\,;\, \Psi, \Lambda)\delta\Lambda  \Big]   \,, \nn 
\end{align}
where we have assumed that $\delta_{\Lambda}\xi^{\mu\nu}_{H}=0$ and have used the linearity of $Q^{\mu\nu}_{ADT}(\xi\,;\, \Psi\,, \delta \Psi)$ on its variable $\delta \Psi$. Here, $\Delta'$ for the $\Omega$-potential denotes the subtraction by the `background contribution'. 
%
%

At the end,  one can see that the thermodynamic volume of the black hole is given by
%
%
\begin{align} \label{volDef}
V = \Big( \int_{\infty} - \int_{\cal B} \Big)   d^{D-2}x_{\mu\nu}~     \Delta'\Big[ \sqrt{-g} \Omega^{\mu\nu} (\xi\,;\, \Psi, \Lambda) \Big] \,,
\end{align}
which can be regarded as the generalization of the known expression of the thermodynamic volume in  Einstein gravity to a generic theory of gravity. 

In Einstein gravity whose Lagrangian is given by ${\cal L} = R -2\Lambda + {\cal L}_{m}(\psi)$, the $\Omega$-potential for the cosmological constant variation satisfies
\[   
 \mathscr{E}^{\mu}_{~\nu}(\xi\,;\, \Psi, \Lambda)\xi^{\nu}\,\delta \Lambda = \nabla_{\nu}\Omega^{\mu\nu} \delta \Lambda= \xi^{\mu}\delta\Lambda\,,
\]
where the $\Omega$-potential reduces to the so-called Killing co-potential $\Omega^{\mu\nu}_{\xi}$ where $\xi^{\mu} = \nabla_{\nu}\Omega^{\mu\nu}_{\xi}$.
Then, the thermodynamic volume of the black hole could be shown to be given by
\begin{equation} \label{vol}
V = \int_{\infty} dx_{\mu\nu} \sqrt{-g}\Big[\Omega^{\mu\nu}_{\xi_{H}} - \Omega^{\mu\nu}_{bg, \, \xi_{H}}\Big] -\int_{\cal B} dx_{\mu\nu} \sqrt{-g}\Omega^{\mu\nu}_{\xi_{H}} \,,
\end{equation}
where $\Omega^{\mu\nu}_{bg}$ denotes the `background contribution'. Note that there is no such contribution from the horizon since the entropy of black holes for a Killing horizon  could be written solely in terms of the Noether potential as was shown by Wald~\cite{Wald:1993nt}. This result is completely matched to those in the literatures~\cite{Kastor:2009wy,Brenna:2015pqa, Kubiznak:2016qmn}. (See, also~\cite{Wu:2016auq} for a covariant phase space approach for a generic theory of gravity.)  As is clear from the construction,  this thermodynamic volume satisfies the extended first law of black holes given in Eq.~(\ref{extFirstLaw}) even for a higher derivative gravity and/or the gravity with various hairy matter fields.

 Now, let us consider a model with a scalar field potential whose  the overall coefficient is given by the cosmological constant. Specifically, consider the scalar potential of  the form $U(\varphi) = 2\Lambda h(\varphi)$. One may note that in this way the scalar field has mass dimension zero and all the self-interacting coupling constant of scalar field are dimensionless. In this case,  the $\Omega$-potential is determined by the following relation
\begin{equation} \label{ScalarPot}
h(\varphi)\xi^{\mu} = \nabla_{\nu}\Omega^{\mu\nu}_{\xi}\,.
\end{equation}
Therefore, at the formal level, the $\Omega$-potential is given by $\Omega^{\mu\nu}_{\xi} \sim  \int dx^{[\nu} \xi^{\mu]}h(\varphi)$. In the following section, the concrete example for this case will be given  and the hairy contribution to the thermodynamic volume of the black holes will be discussed.

Some comments are in order. 
To achieve the consistent thermodynamic interpretation for varying cosmological constant $\Lambda$, it would be essential to assume that  the variation with respect to the cosmological constant, denoted as $\delta_{\Lambda}$, and  the variation  with respect to the other parameters, denoted as $\delta$, commute:
\[   
\delta \delta_{\Lambda} - \delta_{\Lambda}\delta =0\,.
\]
Let us assume that this `integrability'  condition  holds with the condition $\delta_{\Lambda}\xi^{\mu}=0$, which corresponds to a specific parametrization of solutions in terms of usual black hole parameters with the cosmological constant. 
Now, let us consider the effects  on the first law   of the variation with respect to the cosmological constant. 
In order to see this, we would like to  relate the ADT current to the variation with respect to the cosmological constant.
First, recall that we have required the condition $\delta_{\Lambda}\xi^{\mu}_{H}=0$ for the horizon Killing vector $\xi_{H} \equiv \xi_{T} +\Omega_{H}\xi_{R}$,  which implies $\delta_{\Lambda}\Omega_{H}=0$. This condition means that we have chosen a path in the solution space, in which the angular velocity is the variable independent of the cosmological constant, $\Lambda$. 

Instead of $\delta_{\Lambda}\xi^{\mu}=0$, one may take the different parametrization of solutions in such a way that 
\begin{equation} \label{}
\delta_{\Lambda}J =0\,, \qquad \delta_{\Lambda} {\cal S}_{BH}=0\,.
\end{equation}
Then, the mass becomes the function of $\Lambda$ and the thermodynamic volume is given by
\begin{equation} \label{}
\delta_{\Lambda} M = - \frac{V}{8\pi G}\delta \Lambda\,.
\end{equation}
In this choice of the parametrization, the thermodynamic volume becomes
\begin{equation} \label{}
V = -  \int_{\infty}dx_{\mu\nu}~\delta_{\Lambda}\Big(\Delta K^{\mu\nu} - 2\xi^{[\mu}_{T}{\cal B}^{\nu]}\Big)\,,
\end{equation}
where ${\cal B}^{\mu} = \int ds \Theta^{\mu}(\delta_{s}\Psi)$.

We would like to note that, in the theory with $U(1)$ gauge fields, the ADT potential for the background configuration, $Q^{\mu\nu}_{ADT}(\xi_{H}\,;\, \Psi_{bg}, \delta_{\Lambda} \Psi_{bg})$, could be improved as the $U(1)$ gauge  invariant quantity. At the naive application of some formulae, the  expression of $Q^{\mu\nu}_{ADT}$  may be dependent on a large gauge transformation.  However, one could improve the ADT potential $Q^{\mu\nu}_{ADT}$ by adding the total derivative terms so that it becomes independent of the large $U(1)$ gauge transformation. Indeed, by using the on-shell condition given in (\ref{CurrentRel}), we can see that the ADT potential for the background configuration could be related to the $\Omega$-potential for the background configuration as 
\begin{align}
	Q^{\mu\nu}_{ADT}(\xi\,;\, \Psi_{bg}, \delta_{\Lambda} \Psi_{bg}) + \nabla_\rho U^{\mu\nu\rho} = -\Omega_{bg}^{\mu\nu}(\xi\,;\, \Psi_{bg}, \Lambda)\delta \Lambda\,,
\end{align}
where $U$-tensor denotes  the totally antisymmetric tensor $U^{\mu\nu\rho}=U^{[\mu\nu\rho]}$. Since the $\Omega$-potential for the background, $\Omega^{\mu\nu}_{bg}$, is gauge invariant by construction, we can adopt $\Omega^{\mu\nu}_{bg}\delta \Lambda$ as the improvement of the ADT potential $Q^{\mu\nu}_{ADT}(\xi\,;\, \Psi_{bg}, \delta_{\Lambda} \Psi_{bg})$ for the large gauge transformation. 
%

\section{Scaling transformation and the Smarr relation}
In this section, by using a simple model, we set up our conventions and explain the scaling transformation approach to the (extended) Smarr relation~\cite{Hyun:2015tia,Ahn:2015uza,Ahn:2015shg,Hyun:2016isn}. Before presenting some details, it would be better to give some comments on our approach.
Basically, our approach utilizes the reduced action formalism for black hole solutions. Therefore, an appropriate ansatz for the class of black holes is chosen to obtain the reduced action. However, the form of the ansatz is not the essential ingredient in our discussion, as it can be readily generalized to include the other class of black hole solutions. Furthermore, even though the concrete computation depends on the kind of asymptotic geometry, our approach is rather universal in the sense that the asymptotic geometry does not affect its generic features. For instance,  in our reduced action formalism, the divergent expression could appear in the intermediate steps for the asymptotic AdS or Lifshitz spacetime unless appropriate counter terms are added, which render the expressions finite in conjunction with the so-called Gibbons-Hawking terms. Finding these counter terms is hard and tedious part of computations as they depend not only on the model but  also on the asymptotic behavior of the geometry under consideration. As far as we are interested in the Smarr relation among various charges of the geometry, we can bypass these difficulty in the determination of counter terms.

In order to present our method succinctly,  we focus on  a specific model, while some concrete examples are relegated to the next section. 
Let us consider the $D$-dimensional static AdS black hole system with the admissible scalar hair. The action with the cosmological constant is taken as
\begin{align} \label{Action}
	I[g,\,\varphi]=\frac{1}{16\pi G}\int d^D x\, \sqrt{-g}\Big[ R-2\Lambda -\frac{1}{2}(\partial\varphi)^2-U(\varphi) \Big]\,,\qquad \Lambda=-\frac{(D-1)(D-2)}{2l^2}\,,
\end{align}
where $l$ denotes the AdS radius.
Imposing spherical, planar or hyperbolic symmetry on the $(D-2)$-dimensional spatial section of black hole geometry,
one may take the generic ansatz for the metric and the scalar hair as follows:
\begin{align} \label{Ansatz}
	ds^{2} &= -e^{2A(r)}f(r)dt^{2} + \frac{dr^{2}}{f(r)} + r^{2}d\Sigma^{2}_{k}\,,\qquad 	d\Sigma_k^2 = d\theta^2+\frac{\sin^2(\sqrt{k}\theta)}{k}d\Sigma_{D-3}^2\,,\\
	\varphi &= \varphi(r)\,,
\end{align}
where $\Sigma_{D-3}$ denotes $(D-3)$-sphere and $\Sigma_{k}$ does $(D-2)$-dimensional space with the constant curvature whose sign  is determined by $k=-1,0,1$. The reduced action, by inserting the above ansatz in the original action, becomes
\begin{equation} \label{}
I_{red}[f,A,\varphi] = \frac{ 1 }{16\pi G}\int d^{D}x L_{red}\,   = \frac{ \Delta t\,\text{Vol}_{\Sigma_{k}} }{16\pi G}\int dr\, \Big(L_{g\,red} +L_{\varphi\,red} \Big)\,,
\end{equation}
where
\begin{align}	
 \Delta t\,\text{Vol}_{\Sigma_{k}} \int dr &= \int d^{D}x\, \bigg(\frac{\sin(\sqrt{k}\theta)}{\sqrt{k}}\bigg)^{D-3} \prod_{n=1}^{D-4}\sin^{n}\phi_{n}\,, \\
	L_{g\,red} &= e^{A} \bigg[ k(D-2)(D-3)r^{D-4} -2\Lambda r^{D-2} - \Big((r^{D-2})'f\Big)' \bigg]\,,\label{Lred}  %
	\\
	L_{\varphi\,red} &= e^A \bigg[ - \frac{1}{2}r^{D-2} f\varphi'{}^2 - r^{D-2} U(\varphi)\bigg] \,.
\end{align}
Here ${}'$ denotes the derivative with respect to the radial coordinate $r$ and  the irrelevant total derivative terms with respect to $r$ are omitted.

Now, we would like to introduce the off-shell scaling transformation in the above reduced action, which has nothing to do with the physical scaling along the radial direction or the scaling with respect to the mass dimension. 
Let us consider the  off-shell `scaling' transformation with arbitrary weight   for each field   in the reduced action as follows:
\begin{align}
	\delta_\sigma f = \sigma\Big(\omega_f f-rf'\Big)\,,\qquad \delta_\sigma e^A = \sigma\Big(\omega_A e^A-r(e^A)'\Big)\,,\qquad \delta_{\sigma}\varphi = \sigma\Big(\omega_\varphi \varphi-r\varphi' \Big)\,,
\end{align}
where $\omega_{f},\omega_{A}$ and $\omega_{\varphi}$ denote the, not-yet-determined, weight of each field $f, A$ and $\varphi$, respectively.  Under the generic field variation for each field $f, A$ and $\varphi$, the reduced action transforms as
\begin{align}
	\delta I_{red} = \frac{1}{16\pi G}\int d^D x\,\Big[ {\cal E}_f \delta f + {\cal E}_A \delta A + {\cal E}_\varphi\delta\varphi +\Theta'(\delta f,\delta\varphi)\Big]\,,
\end{align}
where $\CE_{\Psi}$ denotes the Euler-Lagrange expression for each field $\Psi$ and $\Theta$ does the surface term under a generic variation. On the other hand, the reduced action transforms under the off-shell scaling transformation of the fields as
\begin{align}
	\delta_\sigma I_{red} = \frac{1}{16\pi G}\int d^D x \Big[ \big(-rL_{red}\big)'+\sum_\omega (\omega+1)L_{[\omega]} \Big]\,,
\end{align}
where $L_{[\omega]}$ denotes the weight $\omega$ part of the reduced action. We suppose that the equations of motion from the reduced action can be reproduced from the EOM of the original action. This requirement  would  be achieved by relaxing our ansatz appropriately.
Note that if the  second term in the right hand side of the above equation vanishes, then the off-shell scaling transformation becomes the true {\it symmetry} of the reduced action~\cite{Banados:2005hm,Hyun:2015tia}. That is to say, if all the terms in the reduced action has a weight $\omega=-1$, the reduced action enjoys a  scaling symmetry at the off-shell level.

For the off-shell scaling transformation,
we may define the  current $C(r)$ as
\begin{align}
	C(r)&\equiv\frac{1}{16\pi G}\int  d\Sigma_k \Big[ \Theta(\delta_\sigma f,\delta_\sigma \varphi)-S \Big]\,,\qquad S\equiv-rL_{red}\,.
\end{align}
Note that, as far as the off-shell scaling transformation is not a  true symmetry of the reduced action,  the current $C$ is not conserved along the radial direction.  Indeed, by a simple computation, one can see that the on-shell value of the current $C$ satisfies
\begin{align}\label{divC}
	C'(r)=\frac{ \text{Vol}_{\Sigma_{k}}}{16\pi G} \sum_{\omega}(\omega+1)\big(L_{g\,red}+L_{\varphi\,red}\big)\Big|_{on-shell}\,.
\end{align}
We can rewrite the on-shell expression of $C(r)$ into the form of
\begin{align} 
	C(r)&=\frac{ \text{Vol}_{\Sigma_{k}}}{16\pi G}\bigg[ \frac{\partial (L_{g\,red}+L_{\varphi\,red})}{\partial f'}\delta_\sigma f + \frac{\partial L_{\varphi\,red}}{\partial \varphi'}\delta_\sigma \varphi + r L_{g\,red}+ r L_{\varphi\,red} \bigg] \nn \\[5pt] 
	&= \frac{ \text{Vol}_{\Sigma_{k}}}{16\pi G}  \bigg[(D-2)r^{D-3}e^A\Big(-\omega_f f +rf'\Big) +r^{D-1}e^Af\varphi'{}^2\bigg]\,,  \label{Cexp}
\end{align}
where the constraint equation for the field $A$, ${\cal E}_A=0$, is used in the last equality.
By using the fact that the position of the event horizon  $r_H$ is determined by $f(r_H)=0$  in our ansatz of black holes, the value of the current $C$ at the event horizon is found to be
\begin{align} \label{Cent}
	C(r_H)=\frac{ \text{Vol}_{\Sigma_{k}}}{16\pi G} (D-2)r_H^{D-2}e^{A(r_H)}f'(r_H)\,,
\end{align}
which could be identified with the product of the temperature and  the entropy  as
\begin{equation} \label{CurrEntropy}
C(r_H)= (D-2)T_H {\cal S}_{BH}\,.
\end{equation}
%
%

By using   the asymptotic fall-off behaviors of  $f\sim -\Lambda r^{2}$ and $e^{A}\sim 1$ for the asymptotic AdS geometry without the scalar hair, one can see that
the divergent behavior of the function $C(r)$ is given by  
\begin{equation} \label{DivAdS}
C(r) \underset{ r\rightarrow\infty }{\sim}   -\Lambda \frac{\text{Vol}_{\Sigma_{k}}}{4\pi G}   \frac{r^{D-1}}{D-1} \,,
\end{equation}
which could be rendered finite by introducing appropriate counter terms at the boundary in the original action in the context of the holographic renormalization~\cite{Skenderis:2002wp}. However, we do not take this route in the following, since the relation among those charges could be obtained without the explicit introduction of these counter terms. 

Now, let us extract the finite expression in the following way. First, one can obtain the conserved charges, $Q^{i}_{\infty}$ at the asymptotic infinity by an appropriate method. It turns out that the finite part of $C(r)$ as $r\rightarrow \infty$ includes pieces proportional to these conserved charges. Then, let us extract these finite pieces from the current and denote the remainder as a counter term of this current  as 
\[   
C_{ct}(r) \equiv C(r) -  \sum_{i}\mu_{i} Q^{i}_{\infty}\,,
\]
where $\mu_{i}$ denote the appropriate numerical numbers and/or chemical potentials. 
For example, 
$C_{ct}(r)$ for static charged AdS  black holes can be written as
\begin{equation} \label{Count1}
C_{ct}(r) = C(r) -  (D-3)M -(D-3)\, \mu Q \,,
\end{equation}
where $M$ and $Q$ are the mass and $U(1)$ charge of black holes under consideration\footnote{$C_{ct}$ may, in part, come from the counter terms and the Gibbons-Hawking boundary terms, which are omitted in  our discussion. 
}. 
Finally, the relation from the scaling transformation given in Eq.~(\ref{divC}) can be rewritten in the integrated form as
\begin{equation} \label{FiniteRel}
C(r)-C_{ct}(r) - C(r_{H}) = \frac{\text{Vol}_{\Sigma_{k}}}{16\pi G} \sum_{\omega}(\omega+1) \int^{r}_{r_{H}}dr\, \big(L_{g\,red}+L_{\varphi\,red}\big)|_{on-shell} -C_{ct}(r)\,,
\end{equation}
which is a relation among finite quantities by construction. In fact, since the left-hand side of the equation (\ref{FiniteRel}) is constant, the right-hand side should be a constant as well.

Note that this relation reduces to the usual Smarr relation when the reduced Lagrangian is composed solely of terms of the weight $\omega=-1$, in which the current is conserved and the counter term $C_{ct}(r)$ vanishes~\cite{Banados:2005hm,Hyun:2015tia}. In the following, we would like to address how to interpret this relation as the extended Smarr relation when the cosmological constant is taken as a thermodynamic variable.

When the scaling transformation is not a symmetry of the reduced action,  the choice of the weight of each field in the reduced action may be indeterminate. This indeterminacy may invoke the impression that the scaling transformation is useless. On the contrary, it turns out that this freedom of the choice is quite useful in our context.  Let us elucidate our strategy. As was alluded before, the cosmological constant could be incorporated as one of the thermodynamic variables in the black hole physics~\cite{Brenna:2015pqa, Kubiznak:2016qmn}.  This exhibits that the cosmological constant may  be taken as the same status in the current $C(r)$ as other charges, which indicates   that the cosmological constant could be pulled out explicitly in our off-shell scaling transformation approach.  Then, one may anticipate that the above relation written in terms of the current $C(r)$ implies the extended Smarr relation  with the cosmological constant as a variable. To see this easily, we would like to choose the weight of each field as follows\footnote{Note that this choice of the weights of various fields is not essential, but convenient. The  result should be independent of the choice.}.
We choose the weights for $f$ and $e^A$ such that the terms coming from Einstein-Hilbert action are invariant under the scaling transformation,
which leads to  $\omega_A=-(D-3)$ and $\omega_f=0$. Under this choice, the reduced action term, $L^{\Lambda}_{g\,red}$, coming from the cosmological constant part is not invariant under the scaling transformation. We also choose the weight of the scalar field in such a way that the scalar kinetic term is invariant under the scaling, which tells us  that $\omega_{\varphi}=0$. 
 Under these assignments of weights for various fields, one can, finally,  set
\begin{equation} \label{ExtSmarr}
C(r_0)-C_{ct}(r_0) - C(r_{H}) =  -\frac{\text{Vol}_{\Sigma_{k}}}{8\pi G}  \int^{r_0}_{r_{H}}dr\, r^{D-2}e^{A} \Big(2\Lambda+U(\varphi) \Big) - C_{ct}(r_{0})\bigg|_{\stackrel{r_0 \rightarrow \infty\quad}{on-shell}}
\,.
\end{equation}

As alluded earlier, the right hand side of the above equation is finite as $r\rightarrow \infty$. In fact, if we set
 the thermodynamic pressure as
$
P= -\frac{1}{8\pi G} \Lambda 
$,
we may define the thermodynamic volume of black holes as
\begin{align} \label{ThermVol}
V=  \text{Vol}_{\Sigma_k} \int^{r_0}_{r_{H}}dr\, r^{D-2}e^{A} \Big(1+ \frac{U(\varphi)}{2\Lambda} \Big) +\frac{4\pi G}{\Lambda}C_{ct}(r_{0})\bigg|_{\stackrel{r_0 \rightarrow \infty\quad}{on-shell}}.
\end{align}
%
Surprisingly, as will be checked later, this volume expression is the same as the one in the previous section.
This tells us that we could treat the conserved charges and the volume on equal footing  from the beginning and we could define $C_{ct}(r)$, accordingly.  

In any case, the Eq.~(\ref{ExtSmarr})  could be interpreted as the extended Smarr relation by using the result given in  Eq.~(\ref{Cent}) and by identifying $C(r_{0})-C_{ct}(r_{0})|_{r_{0}\rightarrow\infty}$ as the mass and charges of static black holes\footnote{The mass and other charges of black holes should be computed in an appropriate method to realize this identification. To obtain charges of black holes, we use the quasi-local covariant method  presented in the previous section.}.  The identification of the thermodynamic volume expression in Eq.~(\ref{ThermVol})  is one of our main results, which gives the explicit form of the volume expression in terms of the metric variables  in the ansatz.  

As mentioned earlier, we have two expressions for the thermodynamic volume, one from the quasi-local ADT formalism and the other from the reduced action one. It would be natural to ask whether these two expressions are consistent.  Before checking the consistency of these two expressions in more complicated examples in the next section, one can easily confirm that,  in the case of asymptotic AdS geometry without the scalar hair, the thermodynamic volume (\ref{ThermVol}) from the scaling method is the same as the one by the quasi-local approach in Eq.~(\ref{vol}).
  This is the first check of the consistency for our scaling approach to the extended Smarr relation.
  

\section{Examples}
In this section, we present concrete computations for several examples which reveal  the power of our approach.  Most of all, these examples show that our approach could incorporate the scalar hairy contribution without  any essential difficulty. 

\subsection{AdS-Shwarzschild black holes}

At first, let us consider the AdS-Schwarzschild  black objects ({\it i.e.} $k=-1,0,1$),  which are described in our ansatz by
\begin{align}
	e^A=1\,,\qquad f=k-\frac{m}{r^{D-3}} -\frac{2\Lambda r^2}{(D-1)(D-2)}\,, \qquad \varphi (r) =0 \,.
\end{align}
The mass expression could be obtained by various methods. 
We  use  the quasi-local ADT method, which is explained in some details in  section 2 (see also~\cite{Hyun:2014sha,Hyun:2015tia}), to obtain the infinitesimal mass expression as
\begin{align}
	\delta M &=\frac{1}{8\pi G} \int d^{D-2}x_{\mu\nu}\sqrt{-g}Q_{ADT}^{\mu\nu}  = \frac{\text{Vol}_{\Sigma_k}}{16\pi G}\Big[ -(D-2)r^{D-3}e^A \delta f \Big] = \frac{\text{Vol}_{\Sigma_k}}{16\pi G} (D-2)\delta m \,.
\end{align}
In this simplest case, by comparing the expression of $C(r)$ in Eq.~(\ref{Cexp})  and the above mass expression, the explicit expression of $C_{ct}$ could be taken by 
\begin{equation} \label{DivC}
C_{ct}(r) =  -\Lambda \frac{\text{Vol}_{\Sigma_{k}}}{16\pi G}  \frac{4 r^{D-1}}{D-1}  \,.
\end{equation}
Collecting the expressions for the current, $C(r),~ C(r_H)$ and $C_{ct}(r)$ in Eq.~(\ref{Cexp}), (\ref{CurrEntropy}) and (\ref{DivC}), respectively, one can see that the relation in Eq.~(\ref{FiniteRel}) or equivalently in Eq.~(\ref{ExtSmarr})  becomes
\[
	(D-3)M -  (D-2)T_H {\cal S} = \Lambda \frac{\text{Vol}_{\Sigma_{k}}}{16\pi G}\bigg[\frac{4 r_{0}^{D-1} }{D-1} -\int^{r_{0}}_{r_{H}} dr\, 4r^{D-2}  \bigg]_{r_{0}\rightarrow\infty}\,,
\]
and  the thermodynamic volume of black holes under consideration is given by
\begin{equation} \label{SimpleCasevol}
V =  \text{Vol}_{\Sigma_{k}}\frac{ r_{H}^{D-1} }{D-1} \,,
\end{equation}
which could also be obtained directly from the volume expression in Eq.~(\ref{ThermVol}). 
By using the conventional choice for the bulk pressure as $P=-\Lambda/8\pi G$, one can reproduce the extended Smarr relation~\cite{Kubiznak:2016qmn} together with the cosmological constant  in the form of
\begin{equation} \label{extSmarr1}
(D-3)M = (D-2) T_H {\cal S}_{BH} -2 P V \,.
\end{equation}

\subsection{Lifshitz black hole}

As another example of extended thermodynamic description presented in previous sections, we consider the Einstein-Maxwell-dilaton model with specific analytic solutions which describe black holes and black branes in $D$-dimensional asymptotically Lifshitz geometry~\cite{Taylor:2008tg,Tarrio:2011de}. The action under consideration could be written as
\begin{align}
	I[g,{\cal A},\phi] &= \frac{1 }{16\pi G}\int d^D x\,\sqrt{-g}\, \bigg[ R -2\Lambda -\frac{1}{2} \left(\partial \phi \right)^{2} -\frac{1}{4} \sum^{N}_{i=1} e^{\lambda_{i} \phi}{\cal F}_{i}^2 \bigg]  \,,
\end{align}
where the cosmological constant is related to the dynamical exponent, $z$, and the scale, $\ell$, of the Lifshitz spacetime as $\Lambda=-\frac{(D+z-2)(D+z-3)}{2l^2}$. In this case, the dilaton  field $\phi$ has nothing to do with scalar hairs under consideration  but is introduced, in conjunction with the Maxwell field, to support the asymptotic Lifshitz geometry. 
By taking the ansatz for metric as in Eq.~(\ref{Ansatz}) and for matter fields as $\phi=\phi(r)$ and ${\cal{A}}_{i}= a_{i} (r) dt,$ the reduced action, up to the total derivative terms, can be written as
\begin{align}
I_{red} \left[ A, f, a_i , \phi \right] &= \frac{\Delta t\,\text{Vol}_{\Sigma_{k}}}{16\pi G} \int dr \,e^{A} \bigg[ k(D-2)(D-3)r^{D-4} -2\Lambda r^{D-2}  -\Big( ( r^{D-2} )' f  \Big)' \nonumber\\
&\qquad\qquad\qquad\qquad\qquad -\frac{1}{2} r^{D-2} f \phi'^{2} + \frac{1}{2}r^{D-2} e^{-2A} \sum_{i=1}^{N}  e^{\lambda_{i} \phi} a_{i}'{}^{2} \bigg] \,.\nonumber
\end{align}
In this case, we focus on the case of $k=0$ or $1$, since there are some unresolved issues for  the $k=-1$ configuration.

To use our off-shell scaling method in the reduced action and to obtain the thermodynamic volume term explicitly, we assign weights as follows:
\begin{align}
	\delta_\sigma f &= -\sigma rf'\,,\qquad \delta_\sigma e^A = -\sigma \Big[ (D-3) e^A + r(e^A)'\Big] \,,\nn \\
	\delta_{\sigma}\phi &= -\sigma r\phi'\,, \qquad \delta_\sigma a_{i} =-\sigma \Big[ (D-3)a_{i}+r a_{i}' \Big] \,. 
\end{align}
This choice of weights is taken in the same spirit with the asymptotic AdS geometry in such a way that the cosmological constant term is the unique one breaking the scaling symmetry in the reduced action, since there is no scalar hair in this case. 
In this choice of weights, one can obtain the current $C(r)$ in the form of 
\begin{align}\label{CurrLif}
	C(r)= \frac{ \text{Vol}_{\Sigma_{k}}}{16\pi G}  \bigg[(D-2)r^{D-2}e^A f' +r^{D-1}e^Af\phi'{}^2 - (D-3) r^{D-2}e^{-A} \sum_{i=1}^{N} e^{\lambda_i\phi}a_{i} a_{i}' \bigg]\,,
\end{align}
which leads, on its on-shell value,   to the following relation 
\begin{align}\label{currRelation}
C(r) - C(r_{H}) &=  - 4  \Lambda  \frac{\text{Vol}_{\Sigma_k}}{16\pi G}\,\int^{r}_{r_{H}} dr\,   e^A  r^{D-2} \Big|_{on-shell} \,.
\end{align}
%
%
While $C(r) - C (r_H) $ is invariant under the gauge transform, each of $C(r)$ or $C(r_H)$ contains gauge dependent terms. For the convenience, we choose a gauge as $a_{i} (r_H) =0$  in the following.
%

In order to represent $C(r)$ in terms of  the conserved charges of the black hole/brane at the asymptotic region, let us recall that the black holes and/or  branes are described by~\cite{Tarrio:2011de,Hyun:2015tia}
\begin{align*}
	f &= \frac{r^{2}}{l^{2}}\left[ 1- \frac{m}{r^{D+z-2}} + k \left(\frac{D-3}{D+z-4}\right)^{2} \frac{l^{2}}{r^{2}} \right] 
	\,,\\
	e^A &= \left( \frac{r}{l} \right)^{z-1} \,,\\
	a_1{}' &= l^{-z} \sqrt{2(D+z-2)(z-1)} \,\mu^{\sqrt{\frac{D-2}{2(z-1)}}}\, r^{D+z-3}\,,\\
	a_N{}' &= l^{1-z} \sqrt{\frac{2k(D-2)(D-3)(z-1)}{D+z-4}} \,\mu^{\frac{D-3}{\sqrt{2(D-2)(z-1)}}}\, r^{D+z-5} \,,\\
	e^\phi &=\mu r^{\sqrt{2(D-2)(z-1)}}\,.
\end{align*}

From the above explicit solutions,  it is straightforward to obtain the values of  temperature and entropy as
\begin{align*}
	T_{H} = \frac{1}{4 \pi} f'(r_{H}) e^{A(r_{H})}  
	\,, \qquad {\cal S} = \frac{\text{Vol}_{\Sigma_{k}} r_H^{D-2}}{4 G}\,.
\end{align*}
Indeed, the simple subtraction gives us the following interesting identification
\begin{align}
C(r_{H})&=\frac{\text{Vol}_{\Sigma_{k}}}{16 \pi G}  (D-2) r_H^{D-2} e^{A(r_{H})} f' (r_H) = (D-2) T_H {\cal S}_{BH}  \,. 
\end{align}

At the spatial infinity,  one may compute the mass of black holes by using the quasi-local formalism and  see that
\begin{align}
	C(r_0)  &=\frac{ \text{Vol}_{\Sigma_{k}}}{16 \pi G} \Big[ (D-2) r_0^{D-2} e^{A(r_{0})} f' (r_0)- (D-3) r_0^{D-2}e^{-A(r_0) } \sum_{i=1}^{N} e^{\lambda_i\phi(r_0)}a_{i}(r_0) a_{i}'(r_0)  \Big] \Big|_{r_0 \rightarrow \infty}\nonumber\\
	&= (D-2-z) M + C_{ct}(r_0) -\frac{ \text{Vol}_{\Sigma_{k}}}{16\pi G}  (D-3) r^{D-2}e^{-A} \sum_{i=1}^{N} e^{\lambda_i\phi}a_{i} a_{i}' \Big|_{r_{0}\rightarrow\infty}\,,
\end{align}
where the counter term, $C_{ct}$, denotes the expression which has been introduced to cancel out divergences.   

Now, let us consider the thermodynamic volume of these Lifshitz black branes. 
On one hand, we can apply our algorithm for the extended Smarr relation by the scaling transformation on the reduced action. By taking the pressure as $P=-\frac{\Lambda}{8\pi G}$,  it is straightforward to see (see also~\cite{Brenna:2015pqa}) 
\begin{align}
	-\frac{\text{Vol}_{\Sigma_{k}}}{16\pi G}  (D-3) r_0^{D-2}e^{-A(r_0)} \sum_{i=1}^{N} e^{\lambda_i\phi(r_0)}a_{i}(r_0) a_{i}'(r_0) \Big|_{r_{0}\rightarrow\infty} = 2(z-1) M - \frac{2(z-1)}{D-3+z} P V\,,
\end{align}
where the thermodynamic volume $V$ is given by 
\begin{align}
 V= -\text{Vol}_{\Sigma_k} \int_{r_{H}} dr~ e^{A} r^{D-2}\,.
\end{align}
Here,  the expression $\int_{r_{H}}$ denotes the indefinite integration and the insertion of the value  at $r=r_{H}$. More concretely,  let us denote the indefinite integration of the function $f(r)$ as $F(r)$. Then, the definite integral between the upper limit $r_{0}$ and the lower one $r_{H}$ becomes
\[   
\int^{r_{0}}_{r_{H}}f(r)dr = F(r_{0}) - F(r_{H})\,.
\]
In the above, the expression $\int_{r_{H}}$ denotes  simply $-F(r_{H})$. Note that the upper limit value $F(r_{0})$ is cancelled out by the counter terms. 
Collecting all the results, we obtain the extended Smarr  relation as 
\begin{align}
(D-4+z) M - (D-2) T_{H} {\cal S} = -\frac{2(D-2)}{D-3+z} P V\,.
\end{align}
On the other hand, we can obtain the volume expression,  through the general result given in Eq.(\ref{vol}), as
\begin{align}
 V= -\text{Vol}_{\Sigma_k} \int_{r_{H}} dr~ e^{A} r^{D-2}\,,
\end{align}
which is completely consistent with the previous expression of the thermodynamic volume from the scaling transformation on the reduced action. One may note that the integration over the radial coordinate, $r$, comes from the integrated expression of the Killing co-potential, $\Omega^{\mu\nu}_{\xi}$.
\subsection{ AdS black holes with scalar hair}
In this section, we consider a  three-dimensional gravity model which admits the analytic solution of three-dimensional scalar hairy black holes~\cite{Henneaux:2002wm}. The Lagrangian is given by
\begin{align}
	I[g,\varphi] = \frac{1}{16\pi G} \int d^{3} x \sqrt{-g} \left[ R-\frac{1}{2} ( \partial {\varphi})^{2} - U(\varphi) \right]\,.
\end{align}
In this model, the value of the scalar potential, $U(\varphi)$,  at the asymptotic infinity  plays the role of the cosmological constant. In our convention, the  concrete form of the scalar potential for a one-parameter, $\nu \geq -1$, is given by 
\begin{align}
	U(\varphi) =  2 \Lambda  \left( \cosh^{6} \frac{ \varphi }{4}+ \nu \sinh^{6} \frac{ \varphi }{4} \right) \,.
\end{align}
The analytic black hole solutions are  presented in~\cite{Henneaux:2002wm}, which can be written in our ansatz as
\begin{align}
	e^{A} = \frac{H}{H+2B}\,, \qquad f= \left(\frac{H+2B}{H+B}\right)^{2}  F \,, \qquad  \tanh  \frac{\varphi}{4} = \sqrt{\frac{B}{H + B}}\,,
\end{align}
where the function $H(r)$ and $F(r)$ are given by
\begin{align}
	H= \frac{1}{2} (r+ \sqrt{r^{2} +4 B r}) \,,  \qquad F= -\Lambda\left( H^{2} - (1+\nu) \Big( 3B^{2} + \frac{2B^{3}}{H} \Big) \right)\,.
\end{align} 
The asymptotic geometry of these black holes corresponds to the AdS space, 
and the event horizon is located at 
\begin{align}
	r_H = B \theta_{\nu}\,,
\end{align}
where the parameter $\nu$-dependent constant $\theta_{\nu}$ is given by 
\begin{align}
	\theta_{\nu} = 2 (1+\nu)^{2/3} \frac{ (1+ i \sqrt{\nu} )^{2/3} - (1 - i \sqrt{\nu} )^{2/3} }{ 2i \sqrt{\nu}  } \,.
\end{align}

The Hawking temperature and the entropy of these hairy black holes are easily read as 
\begin{align}
T_{H} = - \frac{3\Lambda}{2\pi } \frac{(1+\nu)}{\theta_{\nu}}  B \,, \qquad {\cal S}_{BH} = \frac{\pi}{2} \theta_{\nu} B \,,
\end{align}
It is also straightforward to reproduce the mass expression of these black holes in Ref.~\cite{Henneaux:2002wm} by the quasi-local ADT method~\cite{Hyun:2014sha}  as 
\begin{align}
	M = - \frac{3\Lambda}{8 G } (1+\nu)  B^{2}\,.
\end{align}
Now we would like to find the thermodynamic volume of black hole by using expressions found by the quasi-local approach in section 2.
By direct computation, we obtain expressions of each term of Eq.~(\ref{exQADT}) as
\begin{align}
\frac{1}{8\pi G}  \int d^{D-2}x_{\mu\nu} \sqrt{-g}  Q^{\mu\nu}_{ADT}(\xi\,;\, \Psi, \delta_{\Lambda}\Psi\,|\,\Lambda) &= \frac{1}{8G}  \big(-r_0^2-2  B r_0 +(5+3\nu) B^{2} \big) \d \Lambda\,,\nn\\
\frac{1}{8\pi G}  \int d^{D-2}x_{\mu\nu} \Big[  \frac{1}{2}\delta_{\Lambda}K^{\mu\nu} -\xi^{[\mu} \Theta^{\nu]} \Big]_{s=0} &= - \frac{ r_0^2 }{8G} \d \Lambda\,,\nn\\
\frac{1}{8\pi G}  \int d^{D-2}x_{\mu\nu}  \int ds~  \xi^{[\mu}\omega^{\nu]}(\delta_{\Lambda}\Psi,\delta_{s}\Psi) &=  \frac{1}{8G}  (	-2 B r_0 +2  B^{2}) \d \Lambda \,.\nn 
\end{align}
Finally, 
 the thermodynamic volume expression in Eq.~(\ref{volDef}) becomes
\begin{align}
V=  3 \pi (1+\nu) B^{2}\,.
\end{align}
%

%

Note that the  volume expression from our off-shell scaling formula given in Eq.~(\ref{ThermVol}) is completely matched to the above result. In this scaling approach, one may note that the  term $C_{ct}$, which is defined in Eq.~(\ref{Count1}),  is given by
\[   
C_{ct}(r) = \frac{1}{8G} r e^{A} ( f' + r f \varphi'^2 ) \,.
\]

Now, let us take the thermodynamic pressure, $P = -\frac{1}{8\pi G} \Lambda $. 
Then, one can check that  the extended Smarr relation is satisfied in the form of 
\begin{align}
	T_{H}  {\cal S}_{BH} -2 PV =0 \,,
\end{align}
which is the three-dimensional version of the extended Smarr relation given in Eq.~(\ref{extSmarr1}). 
As a final check, by using the given explicit expression of the thermodynamic quantities, it is also straightforward to confirm that the extended first law is satisfied in the form of 
\begin{equation} \label{}
\tilde{\delta}M - T_{H}\tilde{\delta}{\cal S}_{BH} = V\tilde{\delta}P\,.
\end{equation}
\section{Conclusion}
In this paper, we have considered  some aspects of the extended thermodynamic relations by treating the cosmological constant as a thermodymaic variable, {\it i.e.} the bulk pressure. This approach is not new and is considered in numerous recent literatures from various perspectives.  Especially, by the Euler's homogeneous scaling argument, the Smarr relation could be derived by the first law or vice versa, even when the cosmological constant is treated as a thermodynamic variable.  The new aspect in our exploration is the inclusion of (scalar) hairy contribution without resorting to the Euler's scaling argument for thermodynamic quantities. Since the (scalar) hairy contribution could be highly non-linear, the Euler's scaling argument is not available and the completely different scaling method is used in the reduced action formulation. 

The main point of our off-shell scaling transformation could be summarized as follows. 
By inserting a specific ansatz in the action, one can obtain the, so-called, reduced action.  By assuming that this reduced action is consistent with  all the equations of motion for the original action in the category of the ansatz, which is the case in our examples, we could use the reduced action to obtain the relation among charges of black holes. Since the Smarr relation connects the quantities at spatial infinity and on the horizon, the scaling symmetry along the radial direction in the  reduced action may be related to the Smarr relation. We have used a specific off-shell scaling transformation  to show that this could be extended to the case even when the transformation is not a symmetry, which is consistently compromised with the treatment of the cosmological constant as a thermodynamic variable. By using the off-shell scaling transformation in the reduced action, we obtain the thermodynamic volume in a somewhat generic form, and then checked its consistency with the extended first law. 

To obtain the extended first law, we have utilized the covariant quasi-local ADT method by extending our previous results. On the contrary to the approach in~\cite{Kastor:2009wy,Kubiznak:2016qmn}, we used the covariant method, which could be applied  rather easily to a generic theory of gravity.  It is verified that the bulk pressure and the thermodynamic volume are consistently incorporated in the Smarr relation and the first law, at least without the scalar hairs in our ansatz.  Though we have shown the consistency in the specific ansatz, we believe that the  scheme of the off-shell transformation  for the Smarr relation could be used in a more generic setup beyond the Einstein gravity and that it could be matched to the extended first law from the quasi-local ADT method, when the scalar hairs do not enter non-linearly. We have also shown that the non-linear scalar hairs lead to the additional contribution to the thermodynamic volume expression in the extended Smarr relation. 

As concrete examples, we have provided several black hole solutions, which could admit scalar hairs.  As anticipated from  the non-linear behavior of the scalar field, the final results for the extended Smarr relation with the non-trivial scalar hair seem to be consistent with the Euler's scaling, and the extended first law is retained in its form even with the scalar hairy contribution, as is presented in Eq.~(\ref{extADTcharge}).  Therefore, it would be very interesting to explore further the Euler's scaling behavior of scalar hairs. For instance, we may ask that the Euler's scaling argument could be extended in some ways  to give us some information about scalar hairs.  More specifically, it would be very interesting to see scaling behavior of scalar hairs in some analytic hairy solutions or to check the extended Smarr relation with the scalar hairy contributions. Another interesting direction is to investigate the interpretation of the extended first law and the Smarr relation in the context of the  AdS/CFT correspondence. It would be very interesting to see how our off-shell scaling or the quasi-local ADT approach are realized in the dual field theory side. In this paper, we have focused on planar black holes for scalar hairy cases, since the existence of scalar hairs in the non-planar black holes is somewhat subtle. Since our approach could be used in these non-planar black holes at least formally, it would also be interesting to explore the relation between the formal derivation of the extended Smarr relation and the existence of scalar hairs in these black holes.

\vskip 1cm
\centerline{\large \bf Acknowledgments}
\vskip0.5cm
{We would like to thank  Byoungjoon Ahn, Kyung Kiu Kim and Miok Park for some discussion.
SH was supported by the National Research Foundation of Korea(NRF) grant 
with the grant number NRF-2016R1D1A1A09917598. SY was supported by the National Research Foundation of Korea(NRF) grant with the grant number NRF-2015R1D1A1A09057057.}
%

\renewcommand{\theequation}{A.\arabic{equation}}
\setcounter{equation}{0}


%
%




\end{document}